# A Deep Learning based Signal Dimension Estimator with Single Snapshot Signal in Phased Array Radar Application

Yugang Ma, Yonghong Zeng, Sumei Sun, Gary Lee, Ernest Kurniawan and Francois Chin Po Shin
*Institute for Infocomm Research, Agency for Science, Technology and Research*
Singapore
{mayg, yhzeng, sunsm, gary_lee, ekurniawan, chinfrancois}@i2r.a-astar.edu.sg

*Abstract*— Signal dimension, defined here as the number of copies with different delays or angular shifts, is a prerequisite for many high-resolution delay estimation and direction-finding algorithms in sensing and communication systems. Thus, correctly estimating signal dimension itself becomes crucial. In this paper, we present a deep learning-based signal dimension estimator (DLSDE) with single-snapshot observation in the example application of phased array radar. Unlike traditional model-based and existing deep learning-based signal dimension estimators relying on eigen-decomposition and information criterion, to which multiple data snapshots would be needed, the proposed DLSDE uses two-dimensional convolutional neural network (2D-CNN) to automatically develop features corresponding to the dimension of the received signal. Our study shows that DLSDE significantly outperforms traditional methods in terms of the successful detection rate and resolution. In a phased array radar with 32 antenna elements, DLSDE improves detection Signal to Noise Ratio (SNR) by >15dB and resolution by > 1°. This makes the proposed method suitable for distinguishing multiple signals that are spatially correlated or have small angular separation. More importantly, our solution operates with a single snapshot signal, which is incompatible with other existing deep learning-based methods.

## I. INTRODUCTION

The signal dimension, namely the number of sources, is widely desired as prior information for further parameter estimations. For example, in the phased array direction finding problem, the well-known high-resolution multiple signal classification (MUSIC) [1] requires knowledge of the number of sources before estimating their direction of arrivals (DoAs). In the aspect of sparse reconstruction, many high-resolution algorithms perform well only when the number of sources is known. For example, the compressive sensing based on the least-absolute shrinkage and selection operator (LASSO). Without knowing the number of sources, the sparse reconstruction could be distorted due to spurious peaks [2].

Existing signal dimension estimators (SDEs) are typically based on Akaike information criteria (AIC) [3,5] and on minimum description length (MDL) [4,5]. The authors in [5] generalized this problem and established the generalized information criteria (GIC) for the estimator. We summarize these works in Section III for comparison purposes.

All of the above traditional signal dimension estimation methods rely on an accurate model of data. However, in practice, accurately modeling data for a wide range of real-world problems is often very difficult. The hypotheses and approximations based on proposers' experiences during modeling may result in missing some features and further lead to insufficient use of the observation data. To this end, deep learning may address these issues for parameter estimation through a data-driven approach. For example, the convolutional neural network (CNN) [6,7] develops features by the algorithm itself through training with a large quantity of observation data. The features extracted by deep learning may differ from that obtained in the traditional methods, but provide more accurate connections between the data and the desired estimate. As a result, the deep learning estimator can, under certain circumstances, outperform the traditional ones. In [12], a one-dimension (1D) CNN deep learning model is designed to estimate the number of independent sources. Due to the requirement of independence, it may not be suitable for estimating the number of signals from the same source but with different delays or reflection directions in communications and radar applications. In [13] and [14], eigenvalues-based deep neural network (DNN) combining regression and classification is proposed for source dimension estimation. Both [13] and [14] rely on eigen-decomposition (EVD) of the covariance matrix of the received signal. In [13], the covariance matrix is obtained through averaging along multiple data snapshots, while in [14], it is obtained using one-bit arcsine law on history data. Note that EVD has high complexity. Also, multiple snapshots of data for the estimation may not be available in applications requiring fast refreshing rate.

In this paper, we propose a DL-based Signal Dimension Estimator, referred to as DLSDE. By making use of 2D-CNN, we convert the signal dimension estimation problem to an analogous image classification problem. Compared to other DL-based approaches, our proposal is suitable for signals with various delay and angular shifting. Further, our proposed method only needs single snapshot data, in which eigen decomposition typically fails, We also show that DLSDE has lower computational complexity, promising a lighter and faster implementation. We also show that DLSDE can estimate signals with tiny unknown parameter difference and outperforms the traditional ones with much higher resolution (sub-1° separation). Thus, our proposed method is suitable for super-resolution algorithms that require *a priori* information on the number of sources.

The rest of the paper is organized as follows. In Section II, the problem is formulated mathematically. In Section III, we review the traditional signal dimension estimators. In Section IV, we present the proposed DLSDE in details including the deep learning neural network architecture and parameter setting. In Section V, the performance of the DLSDE is evaluated, through Monte Carlo simulations, in terms of successful detection rate versus SNR and the angular resolution, and compared against the traditional estimators introduced in Section III. Finally, this paper is concluded in Section VI.

## II. PROBLEM FORMULATION

We consider a received data sequence consisting of $N$ observations, which can be $N$ time samples of a signal frame or from $N$ sensors. The signal reaches the receiver through $K$ unknown paths. Accordingly, this single snapshot received data can be expressed as

$$\mathbf{r} = \sum_{k=1}^{K} a_k \mathbf{b}(x_k) + \boldsymbol{\varepsilon} \quad (1)$$

where $\mathbf{r} \in \mathbb{C}^{N\times 1}$, $a_k$ and $x_k$ are the magnitude and the unknown key parameter of the $k^{th}$ signal, respectively, and $\boldsymbol{\varepsilon} \in \mathbb{C}^{N\times 1}$ is the (complex-valued) additive white Gaussian noise (AWGN) vector, and

$$\mathbf{b}(x) = [\gamma_1(x), \gamma_2(x), \cdots, \gamma_N(x)]^T \quad (2)$$

where $(\cdot)^T$ denotes the transpose operation, and $\gamma_n(x)$ is a complex transfer function with respect to the key parameter, $x$, in the $n^{th}$ observation. In this paper, we take the phased array radar application as an illustrative example. In a Uniformly Linear Array (ULA), $x$ denotes the direction of the arrival (DoA) of the incident plane wave from a far-field source, and

$$\gamma_i(x) = e^{j2\pi f_c \frac{(i-1)d}{c}\sin(x)}, i = 1 \text{ to } N \quad (3)$$

where $f_c$ is the signal carrier frequency, $d$ is the spacing between two adjacent antenna elements (typically set as half wavelength of the signal, $d = \frac{c}{2f_c}$), and $c$ is the speed of light.

While $K$ is unknown *a priori* given the single-snapshot observation $\mathbf{r}$, it can be estimated with signal dimension estimators. The goal of this paper is to estimate $K$ reliably using a deep learning algorithm under such a single-snapshot setting.

## III. REVIEW OF MODEL-BASED SIGNAL DIMENSION ESTIMATORS

For comparison purposes, the existing signal dimension estimators are reviewed in this section. The most typical estimators are based on AIC [3], MDL [4], and their variations [5,8-10]. All of them are based on the eigenvalues of the covariance matrix of the received signal. However, in the single-snapshot setting, the empirical covariance matrix would be rank-deficient (specifically, rank-1), making eigen-decomposition methods not applicable. Instead, by exploiting the regularity and spatial equivariance of the ULA geometry, we can obtain a smaller-sized empirical covariance matrix through spatial smoothing [1,2]. We first form the Hankel matrix based on the observation $\mathbf{r}$, i.e.,

$$\boldsymbol{\Phi}_s = \begin{bmatrix} r(1) & r(2) & \cdots & r(M) \\ r(2) & r(3) & \cdots & r(M+1) \\ & & \vdots & \\ r(N-M+1) & r(N-M+2) & \cdots & r(N) \end{bmatrix} \quad (4)$$

where $r(i)$ denotes the $i^{th}$ element of the received data vector $\mathbf{r}$. The empirical estimate of the covariance matrix with spatial smoothing can be expressed as

$$\widehat{\mathbf{R}}_s = \frac{1}{M} \boldsymbol{\Phi}_s \boldsymbol{\Phi}_s^H . \quad (5)$$

With the spatial smoothing operation, the size of the empirical covariance matrix obtained shrinks to $(N - M + 1) \times (N - M + 1)$, where $M$ is the length of the spatial smoothing. We define $\lambda_i$ as the $i^{th}$ largest eigenvalue of $\widehat{\mathbf{R}}_s$. Let $N' = N - M + 1$. It is evident that $K < N'$. The AIC-based estimator [3,5] for the number of sources is

$$\widehat{K}_{\text{AIC}} = \underset{0 \leq k < N'}{\operatorname{argmin}} \left( M(N'-1)\ln\left(\frac{\frac{1}{N'-k}\sum_{i=k+1}^{N'}\lambda_i}{\left(\prod_{i=k+1}^{N'}\lambda_i\right)^{\frac{1}{N'-k}}}\right) + k(2N'-k) \right), \quad (6)$$

and MDL-based estimator [4,5] for the number of sources is

$$\widehat{K}_{\text{MDL}} = \underset{0 \leq k < N'}{\operatorname{argmin}} \left( M(N'-k)\ln\left(\frac{\frac{1}{N'-k}\sum_{i=k+1}^{N'}\lambda_i}{\left(\prod_{i=k+1}^{N'}\lambda_i\right)^{\frac{1}{N'-k}}}\right) + \frac{1}{2}\left(k\left(2N'-k\right)+1\right)\ln(M) \right). \quad (7)$$

A GIC-based estimator for the number of sources is summarized in [5] as

$$\widehat{K}_{\text{GIC}} = \underset{0 \leq k < N'}{\operatorname{argmin}} \left( M(N'-k)\ln\left( \frac{\frac{1}{N'-k}\sum_{i=k+1}^{N'} \lambda_i}{\left(\prod_{i=k+1}^{N'} \lambda_i\right)^{\frac{1}{N'-k}}} \right) + \alpha(M)(2N'-k)k \right) \quad (8)$$

TABLE 1 BENCHMARK OF SIGNAL DIMENSION ESTIMATORS

|  | **DLSDE** | **[12]** | **[13]** | **[14]** | **AIC/ MDL/ GIC** |
|---|---|---|---|---|---|
| AI model type | 2D-CNN | 1D-CNN | DNN | DNN | NA |
| Signal type | same signal with varying delays and angular shifts | Independent signals | same signal with varying delays and angular shifts | same signal with varying delays and angular shifts | same signal with varying delays and angular shifts or independent signals |
| Based on EVD | No | No | Yes | Yes | Yes |
| Data Snapshots | Single | Multiple | Multiple | Multiple | Single |
| Complexity* | $O(N^2)$ | $O(N)$ | $O(N^3)$ | $O(N^3)$ | $O((N-M+1)^3)$ |

*: kernel and output sizes are fixed.

where $\lambda_i$ is the $i^{\text{th}}$ eigenvalue of $\widehat{\mathbf{R}}_s$ and $\lim_{M\to\infty} \alpha(M) = \infty$ and $\lim_{M\to\infty} \frac{\alpha(M)}{M} = 0$ according to [5]. This was verified in [11] by setting $\alpha(M) = \log_2(M) - 1$. In this paper, we set $\alpha(M) = \sqrt{M}$, which also meets the above criteria on $\alpha(M)$ and simple.

There are other variations to the AIC and MDL [8-10]. However, all such methods can only give an accurate estimation when the sample size is very large [8-10]. For radar applications with highly dynamic scenes, we may only have access to a single snapshot of data within the coherence time window. This leads to a limited sample size, which in turn makes the empirical estimate of the covariance matrix (5), and thereby the corresponding traditional signal dimension estimators, unreliable. This motivates us to adopt a deep learning approach to overcome the challenges of the single-snapshot regime.

IV. DEEP LEARNING BASED SIGNAL DIMENSION ESTIMATOR

Let us treat the estimate of covariance matrix of the received signal (as in (5)) as a 2-dimensional (2D) image; thus, the signal dimension estimation could be seen as a pattern recognition problem, where the task is to recognize the "image" patterns corresponding to different signal dimension values (SDVs). Since the number of possible SDVs is finite, it is straightforward to further pose this as a classification problem by assigning a class to each possible SDV. In this section, we propose the use of deep learning classification architectures for the SDE problem. Let the maximum possible number of sources be $G$; then our task is a $G$-class classification. A $W$-layer neural network solving this problem can be generally represented as

$$\widehat{K} = \underset{1 \leq k \leq G}{\arg\max} f_W\left(f_{W-1}(\cdots f_1(\mathbf{R}))\right) \quad (9)$$

where $f_1$ to $f_W$ are the transfer functions corresponding to the $W$ layers of the neural network, respectively. $f_1$ is the input layer, $f_W$ is the output layer, and the other $W - 2$ layers are the hidden layers. We set $f_W$ to have $G$ outputs, which serve as logits to the classifier—the maximum output is the estimated class, $\widehat{K}$, namely the estimated SDV. Compared to most other types of neural networks, the convolutional neural network (CNN) is more efficient in feature extraction, particularly for 2D inputs (e.g., images). We use 2D-CNN in this paper. In detail, we adopt a 5-layer ($W = 5$) deep learning model consisting of three convolutional layers and two fully-connected layers. The input data of the 2D-CNN is formatted as below,

$$\mathbf{R} = \mathbf{r}\mathbf{r}^H. \quad (10)$$

Note that we do not use $\widehat{\mathbf{R}}_s$ defined in equation (5) as the input of the 2D-CNN, because the deep learning approach does not rely on eigen-decomposition of the covariance matrix. The estimation of the covariance matrix needs spatial smoothing, which sacrifices the matrix size, and hence degrades the detection resolution. In this paper, we set the number of samples, $N = 32$. The model structure and parameter setting of DLSDE are shown in Figure 1. The input layer of the 2D-CNN is a 2D convolutional layer with 2-channel $32 \times 32$ image inputs corresponding to the real and imaginary parts of $\mathbf{R}$, respectively. The optimizer used is the adaptive moment estimation (Adam), and the criterion is the cross-entropy loss. For training the DLSDE model, we generate 200,000 independent samples of $\mathbf{R}$. CDL channel model is followed for receiver signal data generation. The incident angles are drawn uniformly at random from the interval $[-10°, 10°]$ with a minimum DoA separation of $0.1°$. We choose to train the model over 10,000 epochs to allow for the model to converge. During the training, the batch size is set to 1024. To help promote faster training, data with high SNR is used, so as to provide sufficient examples to learn the true relationship between the waveform representation in (10) and the signal dimension; hence, the signal SNR is set to 30 dB. Although data with varying SNR values may be more comprehensive, we have empirically found that such a dataset leads to poor stability on training (which could largely be due to inability to fit the high-noise examples), and that better results are obtained with a fixed high SNR. Nevertheless, we note that while this is true for our proposed architecture, there may be other neural architectures that could better benefit from a wider range of SNR in the dataset, for which we leave the exploration of alternative methods to future work After training, the DLSDE is ready for use.

Compared to the existing deep learning-based signal dimension estimator, our method, DLSDE, does not require multiple data snapshots. It also does not need EVD. This results in a simple and fast solution. As for the computational complexity, approaches in [13],[14] and all model-based approaches rely on EVD, which has a high order of computational complexity $O(N^3)$. For our proposed deep learning approach, because of weight sharing by convolutional layers, CNN is more efficient, in the sense of number of parameters, than the fully-connected DNN model for large-scale data [15]. Table 1 summarizes the benchmark of our proposed and the existing signal dimension estimators.

## V. Numerical Study

In this section, we evaluate the performance of DLSDE with benchmarks against the existing signal dimension estimators. Since our focus is on the single snapshot phased array radar echo signal, the existing deep learning-based solution, which requires multiple data snapshots, is not applicable. The independent signal separation is also out of the current scope. Thus, we only compare DLSDE to model-based traditional estimators for estimating SDV in the numerical study.

### A. Signal Dimension Estimation

For each test item, all compared estimators are provided with 20,000 single-snapshot received data samples, which are independent of the DLSDE training dataset. In Case 1, each received data sample, **r**, includes a random number of sources, from 1 to 4, and their DoAs are uniformly random, distributed in the interval $[-10°, 10°]$ with a minimum DoA separation of $0.1°$. Each individual signal power is randomly set in range $[0,10]$ dB, and 0 dB signal power is the reference for SNR setting. In Case 2, the DoA interval is constrained to $[-5°, 5°]$ instead, while all other parameters remain the same as in Case 1. Estimations are carried out with every single-snapshot received data sample and the successful detection rate—defined as the ratio of the number of correctly estimating the signal dimension to the quantity of total estimations—is computed over 20,000 samples, respectively. For AIC, MDL and GIC based estimators, the spatial smoothing parameter $M$ is set as 17 so as to maximize the detectable signal dimension range. Note that $N - M + 1$ should be larger than the maximum signal dimensions. On the other hand, the size of sliding window, $M$, should also be larger than the maximum signal dimensions. Otherwise, the signal space cannot be represented.

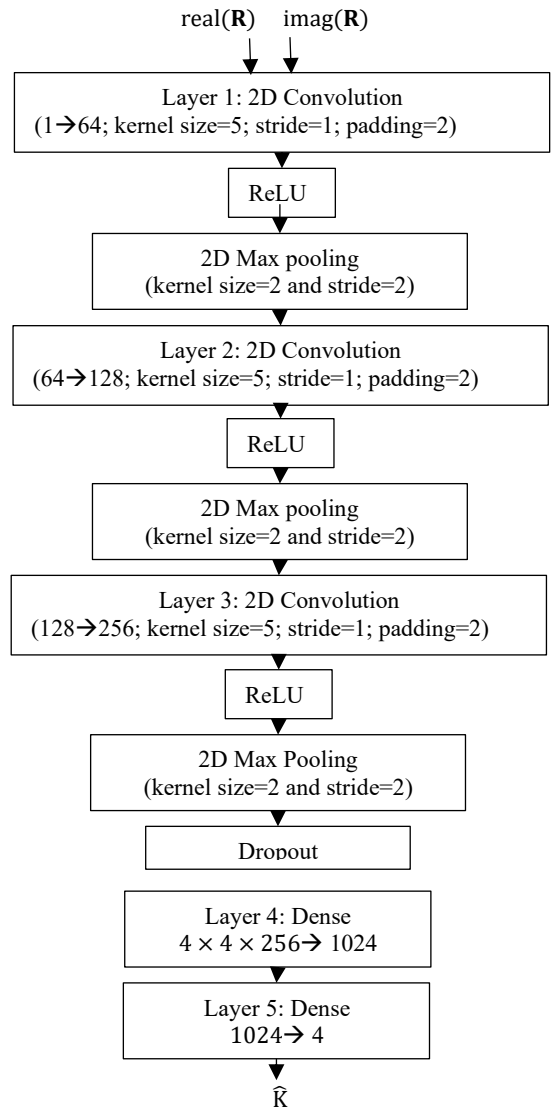

Figure 1. 2D-CNN Model Architecture of DLSDE

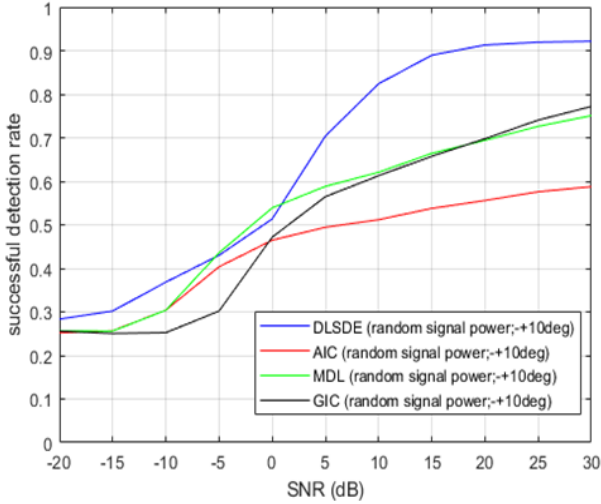

Figure 2. Comparison of the successful detection rate vs SNR ($N = 32$; The number of sources is uniformly random from 1 to 4; minimum DoA separation 0.1°. Individual signal powers are random from 0 to 10 dB; DoAs are uniform random samples from the range $[-10°, 10°]$ )

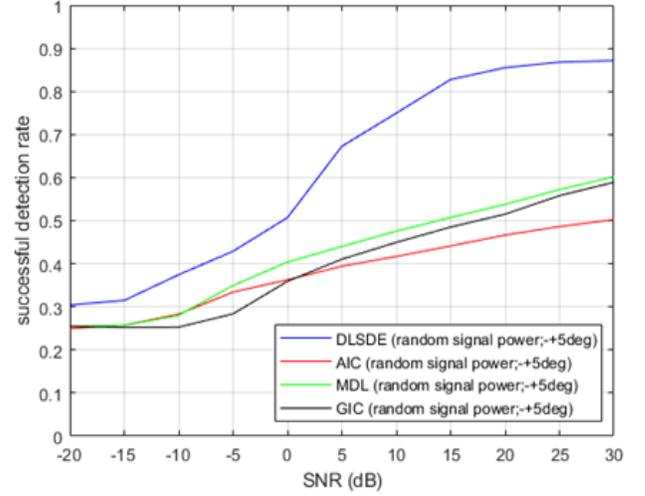

Figure 3. Comparison of the successful detection rate vs SNR ($N = 32$; The number of sources is random from 1 to 4; minimum DoA separation 0.1°; Individual signal powers are random from 0 to 10dB; DoAs are random in a narrower range $[-5°, 5°]$ )

Figure 2 shows the comparison result of the successful detection rate for the signal dimension of the proposed DLSDE and the traditional estimators in Case 1. We can see that DLSDE significantly outperforms the benchmarked ones when SNR > 2 dB. At lower SNR, the DLSDE is comparable to the MDL-based estimator, which is the best-performing method among existing estimators in that SNR regime. Among the traditional estimators, the GIC-based estimator performs better than MDL and AIC estimators in the high SNR regime, i.e., above 20 dB; however, in the lower SNR range, GIC-based estimator performs the worst. On the other hand, our proposed DLSDE performs well across the different SNR regimes, despite training only on high SNR training examples. Further, we note that the proposed DLSDE reaches 90% successful detection rate when SNR is close to 15 dB, whereas conventional estimators only achieve <80% successful detection rate even when SNR = 30dB. In other words, we highlight that the proposed DLSDE improves the detection SNR by at least 15dB to achieve 90% accuracy.

Figure 3 shows the comparison result in Case 2. In this case, having narrower DoA interval, DLSDE's performance gain is more pronounced. Notably, our proposed method reports the highest successful detection rate across all SNRs, outperforming the other conventional estimators. Note that with a narrower DoA interval, there will be smaller bearing angle differences among signals. Our results reflect DLSDE's ability to handle high-resolution scenarios.

### B. Detection resolution

To quantify the performance gain in distinguishing signals with small spatial separation, we evaluate the resolution of the signal dimension estimators. In this paper, the resolution is defined as the minimum unknown parameter difference, $\Delta x$, that can be distinguished by the signal dimension estimators between two signals. We show how the probability of successfully detecting 2 signals changes along with varying $\Delta x$. In this test, two signals have equal power. Their DoAs are random in the interval $[-10°, 10°]$, but always have $\Delta x$ spacing. SNR is fixed to 15 dB for ease of comparison. Figure 4 shows the probability of successful detection versus $\Delta x$ when the powers of two signals are equal. We can see that the DLSDE has much higher resolution than the benchmarked ones, as noted by the remarkably higher detection rate at smaller $\Delta x$. In other words, DLSDE can distinguish multiple signals with much smaller unknown parameter differences (DoA separation in this paper) than the existing ones in the same conditions (SNR=15 dB, N=32 in this paper). Specifically, as reported in Figure 4, DLSDE can detect signals with sub-degree ( 0.8°) resolution with 100% successful detection rate, when SNR is 15 dB. In the same condition, the benchmarked existing signal dimension estimators can only detect multiple signals with resolution > 1.5° . Among the existing estimators, only the GIC-based estimator can achieve 100% successful detection rate for signals with 2.2° resolution when SNR=15 dB. This means DLSDE improves the detection resolution by > 1°.

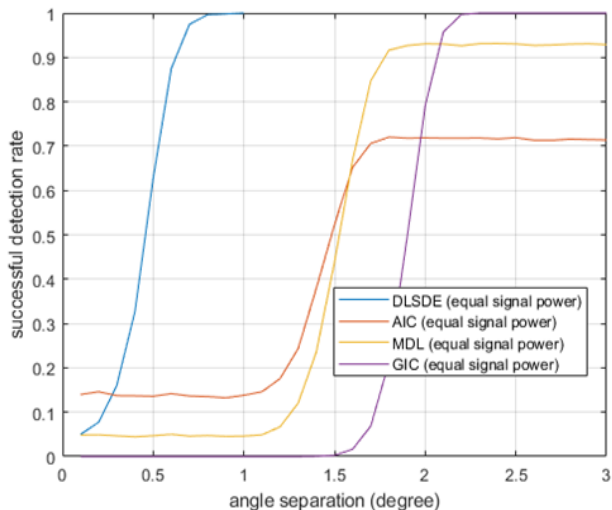

Figure 4. Comparison of the successful detection rate vs angle separation ($N=32$; SNR=10dB; The number of sources is 2; DoAs are random in range $[-10°, 10°]$; SNR=15dB; Two signal powers are equal)

It is worth noting that, in our experiments, the convergence of DLSDE training becomes slow when the maximum signal dimension and the range of the unknown parameter are large. This implies that getting a well-trained DLSDE across a wide range of signal dimensions and/or wide range of unknown parameter values is difficult. Fortunately, in many practical scenarios, the maximum signal dimension in the DoA detection is small. For example, in the phased array frequency modulation continuous wave (FMCW) radar for autonomous driving, DoA detection is done on a given point in the range-Doppler map with detected range and Doppler speed. The signal dimension corresponds to the number of targets with the same range and speed but with different angles, which would be very small in general. DLSDE is especially valuable due to its applicability to high-resolution scenarios. It can distinguish multiple objects in a small bearing angle range (like "zoom-in") for phased array DoA estimation. This is desired in long range radar (LRR) with high-gain transmitter antennas covering a narrow angle range. In this case, the DoA estimation is only conducted within a small field of view (FoV). Nevertheless, we also note the value of DLSDE for high signal dimensions and a wide changing range, which is an area for future work.

## VI. Conclusion

This paper introduces DLSDE, a novel deep learning-based signal dimension estimator using single-snapshot data, which is demonstrated for phased array radar applications. DLSDE outperforms traditional methods, exhibiting a higher success rate and improved detection resolution at equivalent SNRs. This makes it particularly effective for estimating the number of sources with high resolution, and especially beneficial in scenarios with narrow fields of view, such as LRR applications employing high-gain transmitter antennas.